# From Cloud-Native to Trust-Native: A Protocol for Verifiable Multi-Agent Systems

Muyang Li
muyang.li@mail.mcgill.ca

**Abstract**

As autonomous agents powered by large language models (LLMs) proliferate in high-stakes domains—from pharmaceuticals to legal workflows—the challenge is no longer just intelligence, but verifiability. Cloud-native infrastructure enabled scalable computation; AI-native systems enabled reasoning; but neither ensures compliance or traceable accountability. We propose *TrustTrack*, a protocol stack that enables trust-native autonomy by embedding structural guarantees—verifiable identity, policy commitments, and tamper-resistant behavioral logs—directly into agent infrastructure. This architecture reframes compliance as a design constraint, not a downstream process, and shifts trust from institutional oversight to cryptographic traceability. Furthermore, we highlight a key economic dimension: much of the value in AI workflows comes not from raw data or output, but from the structural labor—policies, logic flows, decision scaffolds—often contributed by domain experts without attribution or reward. *TrustTrack* lays the groundwork for compensating this labor via verifiable provenance, enabling fairer participation in an emerging open agency economy. We demonstrate use cases in pharmaceutical R&D and cross-border legal collaboration, and argue that the Cloud → AI → Agent → Trust transition defines the next architectural layer for autonomous systems.

## 1. Introduction

Over the past two decades, digital infrastructure has undergone three architectural shifts. First, cloud computing abstracted away physical hardware, enabling elastic storage and scalable compute [1]. Second, AI systems leveraged this infrastructure to train increasingly powerful models capable of perception and reasoning [2], [3]. Today, we are witnessing a third transition: the rise of autonomous agents—systems that not only infer, but also act, collaborate, and adapt with minimal human supervision [4], [5].

This evolution, however, introduces a growing asymmetry between agency and accountability. As agentic systems proliferate across domains such as pharmaceuticals, finance, and law, their operations become increasingly opaque, untraceable, and jurisdictionally fragmented. Existing compliance frameworks—built for centralized, human-led systems—fail to scale in these decentralized, machine-native environments.

We argue that blockchain is not merely a financial instrument or secure ledger [6]–[8], but a structural substrate for trust-native infrastructure. Just as cloud enabled scalable intelligence, and intelligence enabled agentic automation, blockchain now enables verifiable autonomy. This paper introduces a new systems paradigm, the *Trust-Native Stack*, which comprises four architectural layers: Cloud → AI → Agent → Blockchain.

We present both the architectural framing and a minimal protocol (*TrustTrack*) for enabling cryptographically verifiable agent behavior, including agent identity, policy commitment, and signed behavioral traces. Our vision repositions compliance as architecture, not oversight—and opens the door for structural attribution and compensation within multi-agent ecosystems.

**Contributions**

This is a vision proposal and protocol blueprint. We contribute:

- **Trust-Native Framing**: We position verifiability as a first-class constraint in intelligent systems, alongside latency or throughput.

- **Protocol Design (*TrustTrack*)**: We introduce a lightweight agent auditing mechanism combining DID-based identity, signed actions, and decentralized policy registration.

- **System Analysis**: We illustrate the protocol's relevance in domains like pharmaceutical R&D and legal workflows, and contrast it with existing logging or observability approaches.

1.1 Motivation

As intelligent agents are embedded deeper into critical decision loops, they gain the capacity to act on behalf of organizations, individuals, or even jurisdictions. Whether drafting regulatory reports, coordinating logistics, or approving transactions, these systems introduce new epistemic and legal risks [15], [16]. Their behaviors may be non-deterministic, their reasoning inaccessible, and their provenance unrecorded.

Yet traditional accountability measures—manual review, access control, human attestation—cannot govern machine-native processes at scale. Instead of auditing after deployment, verifiability must be built into the system itself. What we need is not external monitoring, but structural compliance: an architecture where actions are self-disclosing, signatures are verifiable, and traces are tamper-evident.

Furthermore, as multi-agent systems grow in complexity, a second layer of opacity emerges—not just in who did what, but in who designed the logic. Much of the value in agentic workflows arises from upstream structural contributions: prompt templates, policy scaffolds, decision trees, or regulatory knowledge encoded by experts. These contributions shape downstream actions but often leave no audit trail and receive no compensation.

## 1.2 Problem Statement

There exists a foundational mismatch between the growing autonomy of AI agents and the lack of verifiable infrastructure to govern them.

Most agents today operate as black-box entities. When they act—triggering decisions, interacting with other agents, or producing high-stakes outputs—there is often no durable, transparent, or cryptographically secure record of who initiated the action, under what constraints, and with what justification. In regulated environments, this undermines compliance, liability, and cross-organizational trust [17], [18].

Moreover, even when agent behavior is logged, it often resides in centralized, non-portable, or unverifiable formats—making auditability dependent on platform control. What is missing is a protocol-level solution: a system that anchors agent identity, behavioral trace, and policy provenance into decentralized ledgers that are persistent, portable, and independently auditable.

We propose blockchain infrastructure—augmented with agent-level DID registration, selective anchoring, and signature-based policy execution—as the foundation of this trust-native architecture [19]–[21]. This design not only supports verifiable action but also paves the way for structural attribution: a foundation upon which future systems can recognize, trace, and reward non-output labor embedded in autonomous workflows.

## 2. From Cloud-Native to Trust-Native: A Structural Evolution in Autonomous Systems

### 2.1 The First Shift: Cloud-Native Infrastructure and the Rise of Intelligence

Cloud computing marked the first structural leap in digital systems. By abstracting hardware and centralizing compute, cloud-native design enabled elastic workloads, API-based architectures, and real-time data exchange. This shift catalyzed the development of intelligent systems—first machine learning models, then large language models—trained at previously impossible scales. However, this also meant that intelligence became centralized and opaque. AI systems thrived in the cloud, but observability and traceability were sacrificed in favor of performance and scalability.

### 2.2 The Second Shift: From Model-Centric AI to Agent-Based Autonomy

As AI matured, the focus moved from isolated models to autonomous agents. These agents act, reason, and coordinate—often without human oversight—across increasingly distributed, multi-party workflows. Cloud-native infrastructure, while flexible, offers no built-in mechanism to trace, verify, or govern these actions.

In this AI-native era, a new class of failure has emerged: agents taking impactful actions without structural accountability. This exposes a critical gap—autonomy is scaling faster than trust.

## 2.3 The Third Shift: Trust-Native Systems as Compliance Infrastructure

We propose a third structural leap: Trust-Native systems. In this model, verifiability is not a bolt-on feature, but a foundational constraint. Each agent is cryptographically identifiable, its behavior traceable, and its policy commitments verifiable.
This transforms compliance from a post-hoc governance problem to a design-time property of the system. Much like fault tolerance or latency, compliance becomes architectural.
Blockchain infrastructure plays a central role. Rather than serving as a financial tool, it functions as a trust substrate that anchors behavior into shared, tamper-resistant ledgers. This shift from performance-first to accountability-first design defines the Trust-Native paradigm.

This layered evolution reframes system design: every new capability introduces a new class of opacity, requiring structural compensation. Table 1 summarizes this progression and highlights how trust-native infrastructure addresses the blind spots of prior architectures.

**Table 1.** Structural Evolution of Autonomous Systems and Corresponding Trust Compensation Mechanisms

| Layer | System Type | New Capability | Structural Weakness | Trust Compensation |
|---|---|---|---|---|
| L1 | Elastic compute & data | State not observable | State not observable | IAM, API monitoring |
| L2 | Scalable reasoning | Reasoning not interpretable | Reasoning not interpretable | Trace logging, RLHF |
| L3 | Action delegation | Actions not traceable | Actions not traceable | PoA, Verifiable Identity |
| L4 | Immutable audit trails | Slow, costly for real-time systems | Slow, costly for real-time systems | Selective anchoring, zkExec, Merkle batching |

## 3. Design Requirements for Trust-Native Agents

Building trust-native agents—systems that can operate autonomously while remaining accountable—requires more than traditional system design. In contrast to performance-optimized architectures, trust-native design treats verifiability as a first-class constraint. Especially in high-stakes, multi-organizational, or jurisdictionally fragmented environments, the following three system-level requirements are foundational.

### 3.1 Verifiable Identity & Policy Commitments

Every agent must possess a unique, cryptographically verifiable identity, anchored via decentralized identifiers (DIDs) or equivalent mechanisms [9], [10]. Identity alone, however, is insufficient. Agents must also publicly commit to operational policies—defined as the normative boundaries of behavior they are authorized to perform. These policies may include:

- Role-specific permissions and constraints
- Domain-specific ethical or regulatory rules
- Delegated authority from upstream actors (e.g., institutions, designers, prompt engineers)

This identity-policy pairing enables downstream attribution and verifiable delegation. When an agent acts, its identity not only reveals *who* acted, but also *under what declared scope*—a prerequisite for structural accountability in machine-native ecosystems.

### 3.2 Behavioral Traceability Across Time

Traditional observability systems rely on passive logging and platform-bound telemetry, which are often opaque, non-portable, and easy to tamper with. Trust-native agents require a more rigorous approach—namely, cryptographically signed, protocol-level traces that record key decisions, inputs, outputs, and policy references over time [11], [12].

This traceability must:

- Persist outside the agent's memory or proprietary black box
- Support independent verification by third parties (e.g., auditors, regulators, counterparties)
- Allow selective disclosure of relevant events, while preserving privacy or efficiency via cryptographic techniques (e.g., zero-knowledge proofs, Merkle proofs)

Importantly, the emphasis is not on full introspection of AI weights or reasoning steps, but rather on externalized action evidence: what the agent did, under which commitments, and with what consequences.

### 3.3 Interoperability in Zero-Trust Contexts

Autonomous agents increasingly operate in multi-agent, multi-organization, and multi-jurisdiction workflows—contexts where mutual trust cannot be assumed. This breaks the assumptions of centralized governance or shared infrastructure.

To participate in such ecosystems, agents must support:

- Interoperable identity and messaging standards (e.g., DIDs, VC formats, decentralized inboxes)
- Standardized validation protocols, enabling third-party verification of behavior without needing platform access
- Audit-portable logs that can be presented across institutional boundaries

In this setting, compliance is no longer an internal process, but a shared architectural protocol that enables coordination among semi-trusted parties. Trust-native agents must be able to prove their compliance posture in real time, without relying on reputation, human oversight, or informal agreements.

Together, these three capabilities—verifiable identity, traceable behavior, and cross-domain interoperability—form the foundation for building autonomous systems that are not just intelligent, but structurally accountable. In the next section, we introduce *TrustTrack*, a minimal protocol that operationalizes these principles in a lightweight, composable way.

## 4. Protocol Proposal

To meet the design requirements outlined above, we propose a minimal, extensible protocol architecture that enables verifiable collaboration among autonomous agents. This protocol consists of three core layers: Agent Identity, Policy Commitment, and Behavior Logging, each designed to support decentralized coordination without requiring central trust.

### 4.1 Agent Identity Layer

Each autonomous agent must possess a verifiable identity that can be resolved across systems. This layer assigns a decentralized identifier (DID) to each agent, bound to cryptographic keys and optionally linked to metadata (e.g., model version, organizational affiliation). This identity serves as the root of trust for all subsequent actions and assertions [9].

- **Key features:** cryptographic authentication, revocation support, compatibility with DID standards (e.g., W3C) [9], [10].
- **Purpose**: establish continuity of agent identity across interactions and allow third parties to independently verify agent provenance.

### 4.2 Policy Commitment Layer

Before participating in a workflow, each agent must disclose its operational policy: a signed, machine-readable document that defines the constraints under which it acts. These may include hard-coded limitations (e.g., data boundaries, jurisdictional rules), dynamic parameters (e.g., rate limits), or role-specific permissions.

- **Why it matters:** Policy declarations create a precondition for accountability. They encode what an agent *intended* to do and form the baseline against which future behavior can be evaluated [17], [18].

- **Anchoring:** Policy commitments are hashed and anchored on-chain to ensure immutability and traceability [11], [12].

### 4.3 Behavior Logging Layer

As the agent acts, it emits structured behavioral logs—timestamped records of inputs, actions, and outputs—signed with its private key. These logs are optionally batched, hashed, and committed to a shared ledger using Merkle trees or lightweight anchoring schemes [11], [12].

- **Design principles**: tamper-resistance, standardization, and interoperability.

- **Benefits:** Enables decentralized audit [21], cross-agent coordination [23], and post-hoc verification of compliance.

### 4.4 Minimal *TrustTrack* Schema

We implement these principles via a minimal schema inspired by blockchain-light audit trails. Each log entry includes:

- Agent DID
- Policy reference hash
- Action type and parameters
- Timestamp and execution context
- Optional references to upstream/downstream agents
- Signature

This schema is designed to be extensible across use cases (e.g., pharmaceutical compliance, legal automation) while preserving interpretability [15], [16] and auditability.

Figure 1 illustrates the core protocol stack, mapping how agent identity, behavioral traceability, and ledger anchoring work together to enable trust-native autonomy.

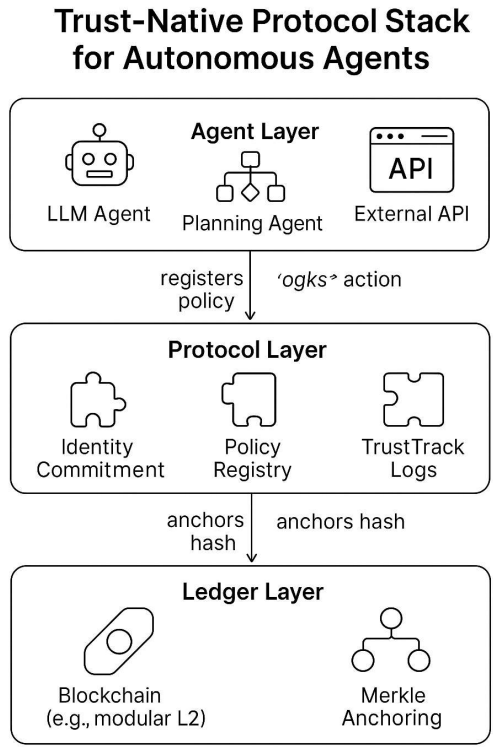

**Figure 1.** Trust-Native Protocol Stack for Autonomous Agents

### 5.4 AI-Native Open Collaboration

As foundation models evolve into autonomous code-generating agents, we are witnessing the emergence of large-scale, multi-agent software ecosystems—where AI systems not only write code, but also review, refactor, and propose changes collaboratively [23]. Much like human developers on GitHub, these agents engage in pull-request workflows, dependency updates, and security patching. But unlike human teams, their provenance, responsibility, and behavior are harder to track.

In such AI-native collaboration, trust-native infrastructure becomes critical. When dozens or hundreds of agents contribute to the same codebase—some fine-tuned for specific domains, others trained by third parties—questions arise: Who generated a given function? Was it reviewed? Did it comply with security policies? Without structured accountability, the open agent economy becomes a liability minefield [17], [18].

A trust-native protocol enables each autonomous contribution to carry a verifiable signature, policy declaration, and behavioral log. Commit histories become cryptographically anchored trails: not just what changed, but who (which agent), under what constraints, and why. This allows for modular trust boundaries: regulators can verify compliance, developers can trace lineage, and institutions can audit AI-generated systems with confidence [21].

Over time, such protocols could underpin a new kind of agent-native software commons—decentralized, composable, and governed by transparent behavioral rules, not just static code licenses [25].

## 5. Illustrative Use Cases

To ground the proposed protocol in real-world settings, we highlight four domains where trust-native agent systems are not only beneficial but increasingly essential: pharmaceutical R&D, cross-jurisdictional legal workflows, smart public infrastructure, and AI-native open collaboration.

Figure 2 outlines the verifiable decision lifecycle of a trust-native agent, detailing how protocol commitments, behavior logging, and audit verification unfold across execution stages.

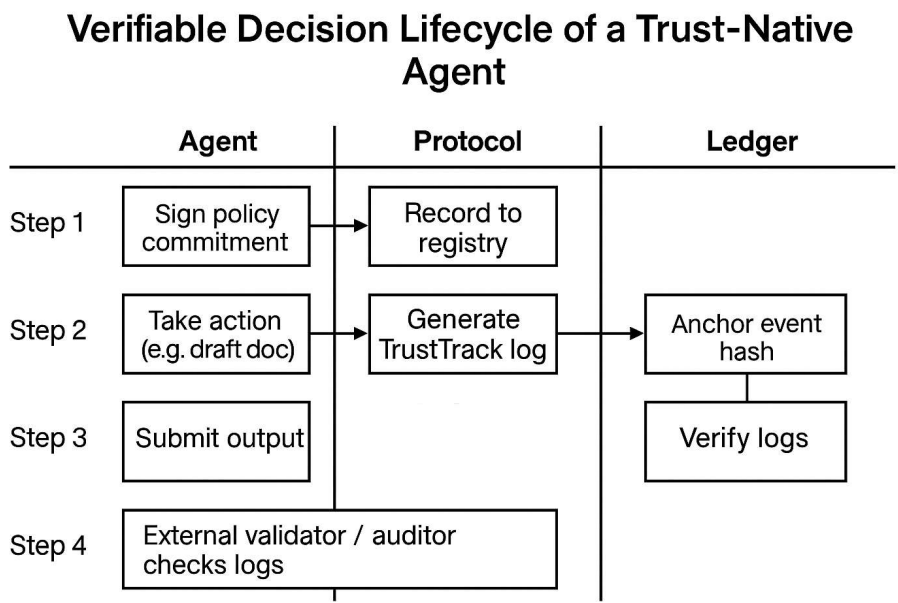

**Figure 2.** Verifiable decision lifecycle of a trust-native agent

•

### 5.1 Pharmaceutical R&D and Regulatory Compliance

Modern pharmaceutical pipelines are becoming increasingly automated and globally distributed. AI agents are now deployed to support:

- Preclinical data synthesis
- Clinical trial documentation
- Adverse event analysis

- Regulatory submission drafting (e.g., IND, NDA, CTD modules)

Each stage demands precise traceability, especially when outputs are submitted to health authorities (e.g., FDA, EMA). With current siloed systems, it remains difficult to establish authorship, model behavior, and intent behind submissions [17], [18].

**How the protocol helps:**

- Agents operating under declared regulatory policies (e.g., GxP compliance) log their actions in verifiable, time-sequenced trails.
- When documentation is reviewed, regulators can cryptographically verify which agent generated what content, under which constraints, and when [21].
- Failures (e.g., hallucinated trial summaries) can be traced to root causes with clear attribution of responsibility.

This reduces the manual QA burden, supports traceable automation, and strengthens institutional trust in AI-assisted drug development workflows.

## 5.2 Cross-Jurisdictional Legal Workflows

In high-stakes legal processes—such as mergers and acquisitions, international arbitration, and data privacy enforcement—multiple agents from different firms and regions often collaborate across conflicting regulatory frameworks.

For instance, a due diligence process in a cross-border M&A may involve:

- AI agents summarizing documents across multiple jurisdictions
- Automated redaction according to region-specific privacy laws
- Legal officers reviewing AI-generated briefs for compliance

**The challenge:** If a redaction error exposes sensitive data, liability is unclear—does it lie with the model vendor, the local legal team, or the orchestrator? [17], [18]

**How the protocol helps:**

- Each agent logs its contributions, bound to a declared policy reflecting relevant legal constraints (e.g., GDPR, CCPA).
- Logs include upstream references, enabling traceability even in multi-agent pipelines [11], [12].

- In the event of an error, firms can cryptographically verify—without relying on a central authority—who took which action and whether it complied with declared constraints [21].

This enables modular, cross-firm collaboration while maintaining accountability and auditability.

### 5.3 Smart Public Infrastructure

In smart cities, critical infrastructure systems—such as traffic control, energy distribution, and public health monitoring—are increasingly delegated to autonomous agents operating across institutional and administrative boundaries [4].

These agents may include:

- A traffic-routing agent deployed by the city
- An energy optimizer from a private grid operator
- A public health AI managing air pollution alerts

In such settings, decisions often interact in real time and impact the public at scale. When failures occur, opaque logic and unverifiable coordination can erode public trust [15], [16].

**How the protocol helps:**

A trust-native framework enables agents to:

- Register cryptographically verifiable identities [9]
- Declare operational policies in advance [10]
- Log actions via tamper-resistant records [11], [12]

This structure allows city agencies, regulators, and independent auditors to reconstruct decision flows during disputes or disruptions [21]. More broadly, it supports a decentralized, interoperable infrastructure ecosystem, where agents can safely negotiate and coordinate actions—without requiring centralized control [24], [25].

### 5.4 AI-Native Open Collaboration

With the rise of autonomous code-generating agents, we are witnessing the formation of large-scale, multi-agent software ecosystems. In these ecosystems, AI agents perform tasks such as:

- Code generation
- Dependency updates

- Peer review and refactoring

- Security patching and compliance enforcement [23]

Much like human developers on GitHub, these agents interact in pull-request workflows and shared repositories. However, unlike humans, their contributions often lack provenance, making it difficult to determine authorship, review status, or policy compliance.

**How the protocol helps:**

- Every autonomous contribution carries a verifiable signature, policy declaration, and behavioral log.

- Cryptographically anchored commit histories clarify what changed, who (which agent) changed it, and under what constraints.

- Institutions and developers can audit AI-generated code with confidence, verifying both intent and compliance [21].

Over time, such infrastructure could underpin a new type of software commons—composable, decentralized, and governed not only by static licenses, but also by dynamic behavioral guarantees [25].

# 6 Related Work

Efforts to enhance the transparency and coordination of AI agents span multiple research domains, including cloud observability, multi-agent systems, and blockchain-based AI auditing. Each addresses part of the accountability challenge, but none provide a unified protocol that integrates agent identity, policy declarations, and behavioral traceability for real-world, cross-organizational agent collaboration.

## 6.1 Cloud Observability Frameworks

Observability frameworks such as OpenTelemetry [26] are widely adopted in cloud-native environments, offering telemetry and tracing tools for distributed microservices. These systems enable service-level logging and diagnostics within trusted organizational boundaries. However, they lack cryptographic guarantees, agent-level identifiers, or any notion of declared operational policy. As such, they are insufficient for scenarios involving decentralized, multi-agent behavior or external regulatory requirements.

## 6.2 Multi-Agent Simulation and Coordination

Platforms like AgentVerse [28] and memory modeling systems such as MEMO [27] focus on agent reasoning, interaction, and planning in simulated environments. These tools are valuable for exploring emergent behavior and cooperative task delegation. Yet, they are largely sandboxed: they do not persist actions in auditable form, nor do they offer verifiable identity or execution constraints. *TrustTrack* complements this space by providing persistent, protocol-driven traceability suitable for deployment in production and regulated settings.

### 6.3 Blockchain-Based AI Auditing

Recent work such as Accountable AI [21] and Decentralized AI [23] proposes blockchain as a tamper-resistant ledger for anchoring AI outputs. These approaches introduce important primitives for auditability and provenance but tend to focus on post-hoc verification of model outputs, without encompassing the full lifecycle of agent behavior. Moreover, identity, policy, and action are typically treated as disjoint elements. *TrustTrack* differs by offering a protocol that integrates all three as part of a unified, composable architecture.

## 7. Discussion & Implications

The transition from cloud-native to trust-native system design represents a fundamental shift in how autonomous systems are built. Instead of applying oversight mechanisms after the fact to black-box agents [15], [16], the trust-native approach incorporates verifiability as a core system constraint. This reframes autonomy as not only a matter of performance and scalability, but also one of persistent accountability and traceable coordination.

In domains such as pharmaceuticals, legal services, and public infrastructure, this model offers a structural pathway to align AI behavior with legal and ethical requirements. By embedding verifiable agent identities [9], operational policies declared in advance [10], and tamper-resistant behavioral records [11], [12], trust-native protocols provide the guarantees needed for compliance, governance, and inter-organizational collaboration. These features are essential in workflows where decisions span multiple jurisdictions or where agents act on behalf of regulated institutions [17], [18], [21].

Beyond compliance, trust-native architecture supports a broader model of socio-technical collaboration. When agents are not only intelligent but also accountable by design, they can interoperate safely across institutions. This opens the door to a modular and transparent form of autonomous coordination, enabling what we describe as an open agency economy. In such a system, models, policies, and behaviors function as composable elements—auditable, portable, and interoperable by default [23], [24].

To situate *TrustTrack* within existing systems, Table 2 provides a comparison with representative frameworks from observability, multi-agent infrastructure, and blockchain-based AI auditing.

**Table 2.** Comparison of *TrustTrack* with representative systems across key dimensions of agent accountability.

| System | Domain | Agent Identity | Policy Declaration | Tamper-Proof Logs | Inter-Agent Support | Regulatory Suitability |
|---|---|---|---|---|---|---|
| OpenTelemetry | Cloud observability | ✗ | ✗ | ✗ | Limited (service-level) | ✗ |
| DeepMind MEMO | Agent memory modeling | ✗ | ✗ | ✗ | ✗ | ✗ |
| AgentVerse | Multi-agent simulation | ✗ | ✗ | ✗ | ✓ (sandboxed only) | ✗ |
| Accountable AI [21] | AI output auditing | ✓ | ✗ | ✓ | ✗ | Partial |
| *TrustTrack* (ours) | Verifiable autonomy infra | ✓ | ✓ | ✓ | ✓ | ✓ |

As the table demonstrates, existing systems each address a subset of the verifiability challenge. Tools like OpenTelemetry provide observability within trusted cloud environments but lack mechanisms for provenance or policy anchoring. Agent simulation platforms such as AgentVerse enable coordination among agents but do not support behavior logging or auditability. Blockchain-based proposals like Accountable AI introduce cryptographic logs but stop short of providing a layered protocol that captures identity, declared intent, and execution context as a unified trace.

*TrustTrack* aims to fill this gap. It offers a minimal and extensible protocol that integrates decentralized identity, machine-readable policies, and cryptographically signed behavioral logs into a single, verifiable structure. This architecture is well-suited for real-world applications where transparency, accountability, and cross-boundary collaboration are non-negotiable.

Several challenges remain. Introducing verifiability increases system complexity and resource demands. Efficient implementation will require advances in lightweight cryptographic primitives, on-chain/off-chain anchoring strategies, and identity lifecycle management. In addition, protocol governance—defining policy standards, maintaining trust registries, and enabling dispute resolution—remains an open area for future development [19], [20], [25].

Despite these open questions, the direction is clear. As autonomous systems grow in capability and institutional importance, trust cannot remain an emergent property. It must be intentionally constructed, encoded into the system architecture itself. *TrustTrack* represents one possible blueprint for such a foundation.

## 8. Conclusion

As autonomous agents move from isolated systems to networked actors in high-stakes workflows, the infrastructure that supports them must evolve. Cloud-native patterns optimized for scalability and modularity [1] are no longer sufficient. What these systems now require is structural trust—built into the fabric of agent identity [9], [10], behavior [11], [12], and coordination [24].

This paper proposes a shift toward trust-native design, where verifiability is treated as a system-level constraint. We outline the core requirements for such systems—identity anchoring, behavioral traceability, and cross-domain interoperability—and present a protocol-based approach grounded in blockchain infrastructure [19], [20]. By anchoring agent actions into shared, tamper-resistant records, we enable a new class of accountable [21], auditable, and collaborative AI-native ecosystems [23].

Trust-native infrastructure is not a niche concern. It is foundational to the safe, legal, and scalable deployment of intelligent systems in fields where consequences matter [17], [18]. As the capabilities of AI agents grow, so must our ability to govern them—not through manual oversight, but through protocol.